\documentclass[12pt]{iopart}
\usepackage{iopams}
\usepackage{setstack}
\usepackage{graphicx}
\usepackage{epsfig}
\usepackage{url}
\usepackage{epstopdf}

\newcommand{\binom}[2]{\left(\!\!\!\begin{array}{c} {#1} \\ {#2}\end{array}\!\!\!\right)}
\newcommand{\eA}{{\epsilon_{A}}}
\newcommand{\eB}{{\epsilon_{B}}}

\newcommand{\eAB}{{\epsilon_{AB}}}

\newcommand{\eTOT}{{\epsilon_{TOT}}}
\newcommand{\nA}{{n_{A}}}
\newcommand{\nB}{{n_{B}}}
\newcommand{\nAB}{{n_{AB}}}
\newcommand{\nTOT}{{n_{TOT}}}
\newcommand{\bA}{{b_{A}}}
\newcommand{\bB}{{b_{B}}}
\newcommand{\bAB}{{b_{AB}}}
\newcommand{\bTOT}{{b_{TOT}}}
\newcommand{\NA}{{N_{A}}}
\newcommand{\NB}{{N_{B}}}

\newcommand{\NAB}{{N_{AB}}}

\newcommand{\NTOT}{{N_{TOT}}}

\begin{document}

\title{Upper limits from counting experiments with multiple pipelines}
\author{Patrick J Sutton}
\address{School of Physics and Astronomy, Cardiff University, Cardiff, United Kingdom, CF24 3AA}
\ead{patrick.sutton@astro.cf.ac.uk} 

\begin{abstract}
In counting experiments, one can set an upper limit on the rate of a Poisson process based on a count of the number of events observed due to the process.  In some experiments, one makes several counts of the number of events, using different instruments, different event detection algorithms, or observations over multiple time intervals.
We demonstrate how to generalize the classical frequentist upper limit calculation to the case where multiple counts of events are made over one or more time intervals using several (not necessarily independent) procedures.  We show how different choices of the rank ordering of possible outcomes in the space of counts correspond to applying different levels of significance to the various measurements.  We propose an ordering that is matched to the sensitivity of the different measurement procedures and show that in typical cases it gives stronger upper limits than other choices.  As an example, we show how this method can be applied to searches for gravitational-wave bursts, where multiple burst-detection algorithms analyse the same data set, and demonstrate how a single combined upper limit can be set on the gravitational-wave burst rate.
\end{abstract}

\pacs{06.20.Dk, %-- measurement and error theory
04.80.Nn} %-- GW detectors and experiments

% ---- If uncommented, forces page break before starting main text.
%\maketitle

\section {Introduction}
\label{sec:intro}

One of the most familiar applications of classical confidence intervals is to the counting experiment, in which one attempts to measure or place a limit on the rate of a physical Poisson process by counting the number of occurrences of the process observed during some period of time.
For example, for a single measurement (a single count of events) with low background and an expected physical rate comparable to or lower than the background, one typically sets an upper limit; {\em i.e.}, a one-sided confidence interval.  Given a count $n$, the upper limit is that value of the physical rate such that the {\em a priori} probability of measuring more than $n$ events in the experiment exceeds some chosen confidence level.

Various issues may complicate the procedure for setting the upper limit.  For example, if the background is large, there is a well-known problem that the upper confidence limit may be the empty set when the observed number of events is much lower than that expected from the background.  Another more subtle issue is that the decision to report an upper limit versus a two-sided confidence interval can, if based on the data, cause undercoverage, rendering the procedure invalid.  
%This is particularly important when the true physical rate is comparable to the sensitivity level of the experiment.  
Techniques for addressing these issues have been presented in the literature, for example, by the Feldman-Cousins technique \cite{FeCo:98} and the loudest event technique \cite{Brady:2004gt,Biswas:2007ni}.  These can also be addressed by Bayesian methods; see for example \cite{Biswas:2007ni,1984NIMPA.228..120H,1985NIMPA.241..236P,PhysRevD.37.1153,PhysRevD.38.3582,PhysRevD.38.3584}.

In this paper we are concerned with a different complication: when more than one count is made of the number of events.  One example of where this situation arises is searches for gravitational-wave (GW) bursts with LIGO and similar detectors \cite{LIGO,VIRGO,GEO}.  In this scenario the GW signals are expected to have amplitudes near the noise floor of the detectors, and the rate of detectable events is expected to be of order the inverse of the observation time or less.  
%Furthermore, the detector background noise exhibits transient fluctuations which sometime occur in coincidence in different detectors, mimicking a GW burst.  
To improve chances of detection, multiple algorithms are used to analyse the data \cite{Ab_etal:04,0264-9381-22-18-S43,S5firstyear}, each producing its own list of candidate GW bursts.  
The event lists produced by these algorithms, however, are not 
completely independent. They will generally show some correlation 
between which foreground events they detect, and may also show 
some correlation between the background noise fluctuations they 
detect. Furthermore, the data set itself typically is not of uniform 
sensitivity.  For example, the longest data-collection run to date 
for the LIGO-GEO-Virgo network lasted more than two years \cite{abbott-2007}.  
Over this time the sensitivity of each of the instruments changed, and   
at any given time during the run, anywhere between 1 and 5 detectors 
may have been operating.
The challenge to the data analyst in such an experiment is this: given multiple counts of events collected from processing several data sets of different sensitivities and  with different algorithms, how does one set a single limit on the physical event rate?  

There are many options.  The simplest is to take the union of all of the event lists and observation time, effectively converting the multiple observations into a single observation, and computing the upper limit using a standard technique.  This approach ignores differences in the quality of the data from the different epochs, and in the algorithms themselves.  Alternatives include discarding results from select data sets or algorithms (presumably the less sensitive ones), again with the aim of reducing the observations to effectively a single count.  These approaches invariably involve loss of information from the experiment.
Intuitively, one expects to be able to set stronger limits if one uses all of the information from the experiment rather than only a subset of the information.

In this article we propose a general formalism for setting classical upper limits on experiments involving multiple {\em pipelines}, where a pipeline denotes the analysis of a single data set by a single algorithm.  We characterize the observational results and the sensitivities of the experiment in terms of logical combinations of pipelines.
We show that various choices such as taking the union of data sets correspond to particular choices of weighting of measurements.  We propose a specific weighting choice based on the efficiencies (sensitivities) of the logical combinations, and show that it gives stronger upper limits than other choices in typical cases.  
Furthermore, the efficiency weighting choice makes use of all of the experiment results, naturally handles correlated measurements, and tends to be robust against occasional background contamination of counts.  

This paper is organized as follows.  In Section~\ref{sec:sp} we 
review how one sets a classical upper limit on the rate of a 
Poisson-distributed process in a counting experiment.  
In Section~\ref{sec:mp} we generalize the single-count procedure 
to the case of multiple counts.  We discuss various choices of 
the weighting to obtain upper limits, including our sensitivity-based 
proposal.  We demonstrate each procedure for the case of a counting 
experiment using two pipelines, with and without background.  
In Section~\ref{sec:data} we demonstrate how the same procedure 
naturally handles multiple data sets.  Section~\ref{sec:summary} 
contains a few brief remarks on the applicability of the method.

\section{Single-Pipeline Case}
\label{sec:sp}

We briefly review how one sets a classical upper limit (a one-sided 
confidence interval) on the rate of a Poisson-distributed process 
via a counting experiment.

Consider an experiment that measures the number of events of a specific 
random process that occurs in a time $T$.  We assume that the foreground 
events occur independently of one another, with a mean rate $\mu$ that is 
unknown {\em a priori}.  We further assume that the experiment has a 
probability $\epsilon$ of successfully detecting (counting) any given 
event.  Finally, we assume that the mean number of background events 
(due to ``noise'' or effects other than the physical effect of interest) 
in time $T$ is $b$.  Then the actual total number of events (foreground 
plus background) that will be counted in a given time $T$ is Poisson 
distributed, as is easily demonstrated.

Let us divide the observation time $T$ into $M$ equal sub-intervals 
of length $T/M$.  In the limit of large $M$, the probability of one 
event being detected in any given sub-interval is $(\epsilon\mu T+b)/M \ll 1$, 
and the probability of more than one event in the same interval is 
negligible.  The probability of detecting a total of $N$ events over 
the full time $T$ is derived from binomial statistics as the 
probability of $N$ ``successes'' in $M$ ``trials''.  Defining 
$\lambda\equiv\mu T$ as the expected mean number of foreground events 
occurring, we have 
\begin{eqnarray}\label{eqn:poisson}
P(N|\epsilon\lambda+b) 
  & = &  \lim_{M\to\infty} \binom{M}{N}
            \left(\frac{\epsilon\lambda+b}{M}\right)^N
            \left(1 - \frac{\epsilon\lambda+b}{M}\right)^{M-N} 
            \nonumber \\ 
  & = &  \frac{(\epsilon\lambda+b)^N}{N!} \e^{-(\epsilon\lambda+b)}  \, .
\end{eqnarray}
This is the familiar Poisson distribution for a process with mean number 
of detected events $\epsilon\lambda+b$.

Given an actual measured number $n$, the Poisson distribution 
(\ref{eqn:poisson}) can be used to set an upper limit on the 
value of $\lambda$, or equivalently on $\mu$.  Heuristically, 
values of $\lambda$ much larger than $(n-b)/\epsilon$ are unlikely 
to produce only $n$ detected events.  More formally, we select 
a confidence level $\alpha\in(0,1)$.  The frequentist upper limit 
$\lambda_\alpha$ at confidence level $\alpha$ given $n$ measured 
events is that value of $\lambda$ at which there is an {\em a priori} 
probability $\alpha$ of measuring more than $n$ events.  Implicitly, 
$\lambda_\alpha$ is given by
\begin{eqnarray}
\alpha
  & = &  \!\! \sum_{N=n+1}^\infty \!\! P(N|\epsilon\lambda_\alpha+b) 
     =   1 - \sum_{N=0}^n P(N|\epsilon\lambda_\alpha+b) \, . \qquad
\end{eqnarray}
We define the cumulative probability $C(n|\epsilon\lambda+b)$ as 
the {\em a priori} probability of detecting $n$ or fewer events: 
\begin{equation}\label{eqn:C1D}
C(n|\epsilon\lambda+b)  \equiv  \sum_{N=0}^n P(N|\epsilon\lambda+b) \, .
\end{equation}
We can write the upper limit formula for $\lambda_\alpha$ as
\begin{equation}\label{eqn:ul}
C(n|\epsilon\lambda_\alpha+b)  = 1 - \alpha \, .
\end{equation}
For example, the 90\% confidence level ($\alpha = 0.9$) upper limit 
for zero observed events ($n=0$) and zero background ($b=0$) is 
%\begin{eqnarray}\label{eqn:spul0}
%0.1 & = & C(0|\epsilon\lambda_{90\%}) = \e^{-\epsilon\lambda_{90\%}} \, , \\
%\lambda_{90\%} & = & \frac{2.30}{\epsilon} \, .
%\end{eqnarray}
\begin{equation}
%\eqalign{0.1 = C(0|\epsilon\lambda_{90\%}) = \e^{-\epsilon\lambda_{90\%}} \, , \\
%\lambda_{90\%} = \frac{2.30}{\epsilon} \, .}\label{eqn:spul0} 
0.1 = C(0|\epsilon\lambda_{90\%}) = \e^{-\epsilon\lambda_{90\%}} \, , \quad
\lambda_{90\%} = \frac{2.30}{\epsilon} \, . \label{eqn:spul0} 
\end{equation}
For $n=1$ observed events the upper limit is higher (weaker):
\begin{equation}
%\eqalign{0.1 = (1+\epsilon\lambda_{90\%})\e^{-\epsilon\lambda_{90\%}} \, , \\
%\lambda_{90\%} = \frac{3.89}{\epsilon} \, .}\label{eqn:spul1}
0.1 = (1+\epsilon\lambda_{90\%})\e^{-\epsilon\lambda_{90\%}} \, , \quad
\lambda_{90\%} = \frac{3.89}{\epsilon} \, . \label{eqn:spul1}
\end{equation}

To be rigorous, one must prove that the upper limit formula (\ref{eqn:ul}) has a coverage of at least $\alpha$.  The coverage is defined as the fraction of measurements in an ensemble of identical experiments for which the derived upper limit is greater than or equal to the true rate $\lambda_\mathrm{true}$.  To be a valid upper limit with confidence level $\alpha$, one must show that $\lambda_\alpha\ge\lambda_\mathrm{true}$ in a fraction $\ge\alpha$ of experiments {\em for any possible value of} $\lambda_\mathrm{true}$.

It is straightforward to prove that the upper limit formula (\ref{eqn:ul}) 
has the coverage $\alpha$.  First, we note two properties \footnote{From 
(\ref{eqn:poisson}), 
$\rmd C(n|\epsilon\lambda+b)/\rmd\lambda = -\epsilon(\epsilon\lambda+b)^n\e^{-\epsilon\lambda-b}/n! < 0$ 
for $\lambda>0$, $b\ge0$.} of $C(n|\epsilon\lambda+b)$:
\begin{eqnarray}
C(n|\epsilon\lambda+b) & > & C(m|\epsilon\lambda+b) \qquad \mathrm{for}~n>m \, \mathrm{;} \quad 
\label{eqn:prop1}  \\
\frac{\rmd C(n|\epsilon\lambda+b)}{\rmd\lambda} & < & 0 \, .\label{eqn:prop2}
\end{eqnarray}
Let us suppose that the true value of the rate is $\lambda_\mathrm{true}$.  
Let $m$ be the largest integer such that $C(m|\epsilon\lambda_\mathrm{true}+b) \le 1-\alpha$.  
By definition of $m$, in a fraction $\ge \alpha$ of experiments the 
measured number of events $n$ will be larger than $m$.  For these cases 
$C(n|\epsilon\lambda_\mathrm{true}+b) > 1 - \alpha$.  Applying the upper 
limit formula (\ref{eqn:ul}) and noting (\ref{eqn:prop2}), we see that in 
these cases the derived upper limit $\lambda_\alpha$ will be greater than 
$\lambda_\mathrm{true}$.  The coverage is thus established.

We should note that one has the freedom to ignore the experimental 
background when computing the upper limit; {\em i.e.}, one may use 
the approximation $b=0$.  Since the background will increase $n$ 
above the value due to the physics of interest, the upper limit 
derived using $b=0$ remains valid (provides minimum coverage), 
though it will be higher than if we had accounted for the background.  
We will use this approximation in some of our worked examples.  

We also note the well-known phenomenon that the classical one-sided 
confidence interval procedure can produce an empty upper limit when the 
number of observed events is much lower than the background.  For 
example, for $n=0$ observed events and $b=3$ the 90\% upper limit is 
the solution of
\begin{equation}
0.1 = C(0|\epsilon\lambda_{90\%}+3) = \e^{-\epsilon\lambda_{90\%}-3} \, .
\end{equation}
This has no solution with $\lambda_{90\%}\ge0$.  Methods for handling 
this issue have been proposed, for example, by Feldman and Cousins 
\cite{FeCo:98}.  In this paper we consider only one-sided confidence 
intervals, and therefore we will restrict ourselves to case where 
$b\lesssim1$.

\section{Multiple-Pipeline Case}
\label{sec:mp}

\subsection{Formulation}
\label{sec:form}

The simplest example of a multiple-pipeline experiment is one in which 
two different methods or ``pipelines'' are used to count events (by 
processing the same data, watching the same sky, {\em etc.}) over the 
same epoch $T$.  (We'll consider the case of {\em disjoint} data sets 
in Section~\ref{sec:data}.)  Denote the pipelines by $A$ and $B$.  Any 
given event may be detected by pipeline $A$ only, by pipeline $B$ only, 
by both $A$ and $B$, or by neither pipeline.  We characterize the 
sensitivity of the experiment by the three numbers $\epsilon_A$, 
$\epsilon_B$, and $\epsilon_{AB}$:

\begin{itemize}
\item[$\eA$:]  The probability that any given foreground event will 
be detected by pipeline $A$ but not detected by pipeline $B$;
\item[$\eB$:]  The probability that any given foreground event will 
be detected by pipeline $B$ but not detected by pipeline $A$;
\item[$\eAB$:]  The probability that any given foreground event will 
be detected by both pipeline $A$ and pipeline $B$.
\end{itemize}

\noindent
We denote the expected background by the three numbers $\bA$, $\bB$, and $\bAB$:

\begin{itemize}
\item[$\bA$:]  The expected number of background events detected by 
pipeline $A$ but not detected by pipeline $B$;
\item[$\bB$:]  The expected number of background events detected by 
pipeline $B$ but not detected by pipeline $A$; and
\item[$\bAB$:] The expected number of background events detected by 
both pipeline $A$ and pipeline $B$.
\end{itemize}

\noindent
Finally, the outcome of the counting experiment is the set of three 
numbers $\nA$, $\nB$, and $\nAB$:

\begin{itemize}
\item[$\nA$:]  The number of events detected by pipeline $A$ but not 
detected by pipeline $B$;
\item[$\nB$:]  The number of events detected by pipeline $B$ but not 
detected by pipeline $A$; and
\item[$\nAB$:]  The number of events detected by both pipeline $A$ 
and pipeline $B$.
\end{itemize}

\medskip
\noindent
To interpret $(\nA,\nB,\nAB)$ in terms of an upper limit on $\lambda$, 
we first need to compute the joint probability 
$P((\nA,\nB,\nAB|\lambda,\eA,\eB,\eAB,\bA,\bB,\bAB)$.  
This is straightforward; repeating the logic of the single-pipeline case, 
it is easy to see that
\begin{eqnarray}\label{eqn:poisson2}
\lefteqn{P(\NA,\NB,\NAB|\lambda,\eA,\eB,\eAB,\bA,\bB,\bAB) } \nonumber \\ 
  & = &  \lim_{M\to\infty} 
             \binom{M}{\NA}
             \binom{M-\NA}{\NB}
             \binom{M-\NA-\NB}{\NAB}
         \nonumber \\
  &   &  \times 
            \left(\frac{\eA\lambda+\bA}{M}\right)^{\NA}
            \left(\frac{\eB\lambda+\bB}{M}\right)^{\NB}
         \nonumber \\
  &   &  \times 
            \left(\frac{\eAB\lambda+\bAB}{M}\right)^{\NAB} 
            \left(1 - \frac{\eTOT\lambda+\bTOT}{M}\right)^{M-\NTOT} 
            \nonumber \\ 
  & = &  P(\NA|\eA\lambda+\bA) P(\NB|\eB\lambda+\bB) 
         P(\NAB|\eAB\lambda+\bAB)   \, . %\nonumber \\
%  &   &  \mbox{}         
\end{eqnarray}
Here we have defined the total number of events detected,
\begin{equation}\label{eqn:ntot}
\NTOT \equiv \NA + \NB + \NAB \, ,
\end{equation}
the total number of events expected from background, 
\begin{equation}\label{eqn:btot}
\bTOT \equiv \bA + \bB + \bAB \, ,
\end{equation}
and the probability of a given foreground event being detected by 
any combination of pipelines,
\begin{equation}\label{eqn:etot}
\eTOT \equiv \eA + \eB + \eAB \, .
\end{equation}
We see that by choosing to characterize the outcome of the experiment by the number 
of events detected by logical combinations of pipelines, 
the joint probability factorizes to the product of single-pipeline probabilities (\ref{eqn:poisson}).
The measurements of $\NA$, $\NB$, and $\NAB$ can therefore be regarded as statistically independent experiments.
This is a key simplification that makes deriving a combined upper limit straightforward. 

%Similarly, for three pipelines $A$, $B$, and $C$, we have 
%\begin{eqnarray}
%\lefteqn{P(\NA,\NB,\NC,\NAB,\NBC,\NCA,\NABC|\eA,\eB,\eC, \ldots} \nonumber \\
%\lefteqn{\quad \ldots \eAB,\eBC,\eCA,\eABC,\lambda)} \nonumber \\
%  & = &  P(\NA|\eA\lambda) P(\NB|\eB\lambda) P(\NC|\eC\lambda)   
%         P(\NAB|\eAB\lambda)
%         \nonumber \\
%  &   &  \times P(\NBC|\eBC\lambda)  P(\NCA|\eCA\lambda) P(\NABC|\eABC\lambda)  ~\qquad
%\end{eqnarray}
%and so forth. 
In the general case of $p$ pipelines, there are $q \equiv 2^{p}-1$ distinct 
combinations by which an event may be detected.  Using the vector notation 
$\vec{N}$, $\vec{\epsilon}$, and $\vec{b}$, where the vector index 
$i\in[1,\ldots,q]$ labels the distinct combinations, we have
\begin{equation}\label{eqn:factorization}
P(\vec{N}|\lambda\vec{\epsilon}+\vec{b})
  =  \prod_{i=1}^q P(N_i|\lambda\epsilon_i+b_i) \, .
\end{equation}

\subsection{Defining an Upper Limit}
\label{sec:mpul}

To set an upper limit we need first to define a cumulative probability 
distribution $C(\vec{n}|\lambda\vec{\epsilon}+\vec{b})$ corresponding to 
(\ref{eqn:factorization}), analogous to (\ref{eqn:C1D}).  Since the space 
of observation $\{\vec{N}\}$ is multi-dimensional, we have a great deal 
of freedom in how we choose to sum over $\{\vec{N}\}$ to define the 
cumulative distribution.  Put another way, we must chose a 
{\em rank ordering} of $\{\vec{N}\}$.  (For an unbiased limit, this must 
be done before the measurement of $\vec{n}$.)

To construct a confidence belt, we choose a one-parameter family of surfaces ${\cal S}(\zeta)$ that foliates the observation space $\{\vec{N}\}$.  This family is chosen so that for every value of the parameter $\zeta$, the surface ${\cal S}(\zeta)$ divides the space $\{\vec{N}\}$ into two regions: a {\em acceptance region} of low number of events (including the origin, and the surface ${\cal S}(\zeta)$ itself), and a {\em rejection region} of high number of events.   
Our choice of the family ${\cal S}(\zeta)$ is arbitrary, except that the outward normal to each surface must have non-negative components everywhere; this is required to prove coverage, as shown below.
As we shall see, our freedom in the choice of the ${\cal S}(\zeta)$ corresponds to how the various pipelines are ``weighted'' in contributing to the upper limit.

Because of the foliation, every point $\vec{N}$ in the observation 
space lies on exactly one such surface, which we refer to as an 
{\em exclusion surface}.  Hence, each point $\vec{N}$ can be associated 
with a single parameter value, $\zeta(\vec{N})$.  This gives us a rank 
ordering of the $\vec{N}$ defining whether a given $\vec{N}'$ contains 
``more,'' the ``same,'' or ``fewer'' events than $\vec{N}''$. The 
family ${\cal S}(\zeta)$ therefore maps the multi-dimensional space 
$\{\vec{N}\}$ to a one-dimensional space.  This allows us to define a 
cumulative probability $C_{\cal S}(\vec{n}|\lambda\vec{\epsilon}+\vec{b})$ 
by
\begin{eqnarray}\label{eqn:C}
C_{\cal S}(\vec{n}|\lambda\vec{\epsilon}+\vec{b}) 
  & \equiv &  \!\!\!\!\! \sum_{\vec{N}|\zeta(\vec{N})\le\zeta(\vec{n})} 
      \!\!\!\!\! P(\vec{N}|\lambda\vec{\epsilon}+\vec{b}) \, , 
\end{eqnarray}
where the sum is taken over all $\vec{N}$ for which $\zeta(\vec{N}) \le \zeta(\vec{n})$; {\em i.e.,} over all $\vec{N}$ that contain as few events or fewer than $\vec{n}$.  

Given a family of exclusion surfaces ${\cal S}(\zeta)$ and a measured number of events $\vec{n}$, we may use the cumulative probability $C_{\cal S}$ to set an upper limit on $\lambda$ in the same way as is done for the single-pipeline case.
Specifically, for a measured number of events $\vec{n}$, the upper limit $\lambda_\alpha$ at confidence level $\alpha$ is 
\begin{eqnarray}\label{eqn:uln}
C_{\cal S}(\vec{n}|\lambda_\alpha\vec{\epsilon}+\vec{b}) 
  =  1 - \alpha \, .
\end{eqnarray}
That is, the upper limit $\lambda_\alpha$ on the rate is that value for which 
in a fraction $\alpha$ of an ensemble of experiments one would measure a number of events that falls in the rejection region of  ${\cal S}(\zeta(\vec{n}))$.
Put another way, the upper limit is the rate for which one should measure ``more'' than $\vec{n}$ events (a value of $\zeta$ larger than $\zeta(\vec{n})$) in a fraction $\alpha$ of an ensemble of experiments.  

We will consider various simple choices of families ${\cal S}(\zeta)$ and their interpretations shortly.  First, however, we prove that the algorithm (\ref{eqn:uln}) has coverage $\alpha$.

\subsection{Coverage}
\label{sec:coverage}

We now prove that the upper limit formula (\ref{eqn:uln}) has a coverage of at least $\alpha$.  The proof follows that for the single-pipeline case in Section~\ref{sec:sp}.  Again, we note two properties of $C_{\cal S}(\vec{n}|\lambda\vec{\epsilon}+\vec{b})$:
\begin{eqnarray}
C_{\cal S}(\vec{n}|\lambda\vec{\epsilon}+\vec{b}) & > & C_{\cal S}(\vec{m}|\lambda\vec{\epsilon}+\vec{b}) \quad\!\! \mathrm{for} \quad\!\! \zeta(\vec{n})>\zeta(\vec{m}) \, \mathrm{;} ~\qquad \mbox{} 
\label{eqn:prop1n}  \\
\frac{\rmd C_{\cal S}(\vec{n}|\lambda\vec{\epsilon}+\vec{b})}{\rmd\lambda} & < & 0 \, . \label{eqn:prop2n}
\end{eqnarray}
(See Appendix~\ref{sec:proof} for the proof of (\ref{eqn:prop2n}).) 
Let us suppose that the true value of the rate is $\lambda_\mathrm{true}$.  
Let $m$ be the vector with nonnegative integer components and with the 
largest value of $\zeta(\vec{m})$ such that 
$C_{\cal S}(\vec{m}|\lambda_\mathrm{true}\vec{\epsilon}+\vec{b}) \le 1-\alpha$.  
By definition of $m$, in a fraction $\ge \alpha$ of experiments the 
measured number of events $\vec{n}$ will have 
$\zeta(\vec{n})>\zeta(\vec{m})$.  For these cases 
$C_{\cal S}(\zeta(\vec{n)}|\lambda_\mathrm{true}\vec{\epsilon}+\vec{b}) > 1 - \alpha$.  
Applying the upper limit formula (\ref{eqn:uln}) and noting 
(\ref{eqn:prop2n}), we see that in these cases the derived upper 
limit $\lambda_\alpha$ will be greater than $\lambda_\mathrm{true}$.  
The coverage is thus established.

As stated before, our choice of exclusion surfaces is arbitrary except 
that the outward normal to the contour must have non-negative components 
everywhere.  This restriction ensures that equation (\ref{eqn:prop2n}) 
is valid, which in turn is required to prove coverage.  As in the 
single-pipeline case, we may chose to ignore the background and use 
$\vec{b}=0$ when computing upper limits.  Since a non-zero background 
contribution will increase the measured $\zeta$ over its zero-background 
value, from (\ref{eqn:uln})-(\ref{eqn:prop2n}) it follows that the limit 
will be higher than that computed accounting for the background, but 
coverage will be maintained.

\subsection{Choosing Exclusion Surfaces}
\label{sec:surf}

We now turn to the question of how to select the family of exclusion surfaces to obtain the strongest limits.  For simplicity, we restrict ourselves henceforth to the simple case of plane surfaces.  In this case, a family of exclusion surfaces is set by choosing the vector $\vec{k}$ that is normal to the planes.  The parameter for the family is then $\zeta(\vec{N}) = \vec{k}\cdot\vec{N}$ (the magnitude of $\vec{k}$ is irrelevant).  For a given observation $\vec{n}$ the upper limit $\lambda_\alpha$ is given by
\begin{equation}\label{eqn:ulk}
C_{\vec{k}}(\vec{n}|\lambda_\alpha\vec{\epsilon}+\vec{b}) 
  \equiv \!\!\!\!\! \sum_{\vec{N}|(\vec{n}-\vec{N})\cdot\vec{k}\ge0} 
      \!\!\!\!\! P(\vec{N}|\lambda_\alpha\vec{\epsilon}+\vec{b}) 
  =  1 - \alpha \, . 
\end{equation}
Note that the sum is taken over all $\vec{N}$ satisfying the condition
\begin{equation}\label{eqn:contour}
(\vec{n}-\vec{N}) \cdot \vec{k} \ge 0 \, .
\end{equation}

We now explore several simple choices of exclusion surfaces with ready physical interpretations:  taking the logical AND or OR combinations of pipelines, and using only the most sensitive pipeline.  We then propose a new choice of exclusion surfaces: $\vec{k}=\vec{\epsilon}$; {\em i.e.}, we weight the measurements by the relative sensitivity of their pipelines.  We show that this efficiency-weighted approach has several advantages over the other choices discussed.  In particular, it gives upper limits that are better than those from the other common choices for most outcomes of the experiment.

\subsubsection{OR combination} 
\label{sec:or}

One obvious way to orient the exclusion surfaces is to set the normal vector $\vec{k}=(1,1,\ldots,1)$.  This choice treats all distinct pipeline combinations equally.  
For a given observation $\vec{n}$ the upper limit on $\lambda$ is then given by (\ref{eqn:uln}) with the sum taken over all $\vec{N}$ satisfying the condition
\begin{equation}
\sum_i N_i \le \sum_i n_i \, ,
\end{equation}
or simply 
\begin{equation}
\NTOT \le \nTOT \, .
\end{equation}
That is, the upper limit depends only on the total number of events detected, regardless of which pipelines or combinations of pipelines detected them.  We see that this choice of exclusion contour is equivalent to setting an upper limit based on a single pipeline which is formed by taking the ``OR'' combination of all events detected by all pipelines or combinations of pipelines.

For example, consider the case of two pipelines $A$ and $B$. Let us assume 
for simplicity that the background is negligible ($\bA,\bB,\bAB\simeq0$).
If no events are detected, the upper limit at confidence level $\alpha=0.9$ 
is given by
\begin{eqnarray}\label{eqn:ulOR0a}
0.1 
  & = &  C_{\vec{k}}((0,0,0)|\lambda_{90\%}\vec{\epsilon}) \nonumber \\
  & = &  \e^{-\eTOT\lambda_{90\%}} \, ,
\end{eqnarray}
where $\eTOT \equiv \eA + \eB + \eAB$.  This has the solution
\begin{equation}\label{eqn:ulOR0b}
\lambda_{90\%} = \frac{2.30}{\eTOT} \, .
\end{equation}
This has the same form as in the single-pipeline case, (\ref{eqn:spul0}), with the replacement $\epsilon\to\eTOT$.

Now consider the case of one event detected (it does not matter whether the lone event is detected by $A$, by $B$, or by both).  The upper limit is given by
\begin{eqnarray}\label{eqn:ulOR1a}
0.1 
  & = &  P((0,0,0)|\lambda_{90\%}\vec{\epsilon}) 
             + P((1,0,0)|\lambda_{90\%}\vec{\epsilon}) 
             \nonumber \\
  &   &    \mbox{} 
             + P((0,1,0)|\lambda_{90\%}\vec{\epsilon}) 
             + P((0,0,1)|\lambda_{90\%}\vec{\epsilon})
             \nonumber \\
  & = &  (1 + \eTOT\lambda_{90\%}) \e^{-\eTOT\lambda_{90\%}} \, ,
\end{eqnarray}
which has the solution
\begin{equation}\label{eqn:ulOR1b}
\lambda_{90\%} = \frac{3.89}{\eTOT} \, .
\end{equation}
This again has the same form as in the single-pipeline case, (\ref{eqn:spul1}), with the replacement $\epsilon\to\eTOT$.  

The OR combination has the advantage that it has the largest efficiency of any combination, since an event is counted if any of the pipelines detect it.  This leads to strong upper limits when no events are detected.  The disadvantage is that the background is also summed over all pipeline combinations, potentially leading to a high false alarm rate and poor limits if any of the pipeline samples are contaminated by background.

\subsubsection{AND combination} 
\label{sec:and}

A ``conservative,'' choice for detecting events is to demand that {\em all\/} pipelines observe an event for it to be counted as a possible signal.  It is easy to see that this is equivalent to choosing contours with normal vector $\vec{k} = (0,\ldots,0,1)$.  The upper limit for observation $\vec{n}$ is then given by (\ref{eqn:uln}) with the sum is taken over all $\vec{N}$ satisfying the condition
\begin{equation}
N_q \le n_{q} \, ,
\end{equation}
where $n_q$ is the number of events detected in coincidence by all pipelines.
Because of the factorization of the joint probability (\ref{eqn:factorization}), 
the upper limit becomes 
\begin{eqnarray}\label{eqn:ulAND}
1 - \alpha 
  & = &  \sum_{N_1=0}^\infty \cdots \sum_{N_{q-1}=0}^\infty 
            \sum_{N_q=0}^{n_q} P(\vec{N}|\lambda_\alpha\vec{\epsilon}+\vec{b}) 
            \nonumber \\
  & = &  \left[\sum_{N_1=0}^\infty P(N_1|\epsilon_1\lambda_\alpha+b_1)\right] 
            \cdots 
            \nonumber \\
  &   &     \times 
            \left[ \sum_{N_{q-1}=0}^\infty P(N_{q-1}|\epsilon_{q-1}\lambda_\alpha+b_{q-1})\right] 
            \nonumber \\
  &   &     \times 
            \sum_{N_q=0}^{n_q} P(N_q|\epsilon_q\lambda_\alpha+b_q) 
            \nonumber \\
  & = &  \sum_{N_q=0}^{n_q} P(N_q|\epsilon_q\lambda_\alpha+b_q)  \, ,
\end{eqnarray}
We see that the upper limit reduces to that for an effective single pipeline formed by taking the AND combination of all pipelines.  This has the same form as in the single-pipeline case, (\ref{eqn:spul0}), with the replacement $\epsilon\to\epsilon_q$.

Consider again the case of two pipelines $A$ and $B$ with low background.  
Suppose we had decided {\em a priori} to compute an AND upper limit.  
If no events were detected by any pipeline, then the 90\% confidence 
upper limit is given by (\ref{eqn:spul0}) with $\epsilon\to\eAB$. 
\begin{equation}\label{eqn:ulAND0b}
\lambda_{90\%} = \frac{2.30}{\eAB} \, .
\end{equation}
Since $\eAB\le\eTOT$, the AND combination gives a weaker limit for 
a given number of measured events.

Now consider the case in which one event is detected.  The limit now 
depends on which pipeline combination detected the event.  If only 
one of the pipelines detected the event, then $n_q=0$, and the 90\% 
confidence upper limit is given by (\ref{eqn:ulAND0b}).  If both 
pipelines detected the event then $n_q=1$ and
\begin{equation}\label{eqn:ulAND1b}
\lambda_{90\%} = \frac{3.89}{\eAB} \, .
\end{equation}
These have the same form as in the single-pipeline case, (\ref{eqn:spul0}), (\ref{eqn:spul1}), with the replacement $\epsilon\to\eAB$.  

The AND combination has the advantage of being the combination least susceptible to background contamination, since an event is only counted if it is detected by all pipelines.  For example, the AND combination is particularly robust if the pipelines have different responses to the background noise.  The disadvantage is that the efficiency is also the lowest of any combination, for the same reason.  In particular, the AND sensitivity is limited by the least sensitive pipeline.

\subsubsection{SINGLE combination} 
\label{sec:single}

Another simple choice for setting the upper limit is to consider only the measurement by the single most sensitive pipeline, and ignoring all of the others.  The most sensitive pipeline is the one with the largest detection efficiency computed when ignoring the other pipelines; {\em e.g.}, for the two-pipeline case it is the larger of $\eA+\eAB$ (for $A$) or $\eB+\eAB$ (for $B$).  The procedure for computing the upper limit in this case is simply to apply (\ref{eqn:ul}).  We note here that it is another special case of the multiple-pipeline procedure.  For example, for two pipelines where $A$ is the more sensitive, the SINGLE limit is equivalent to choosing 
\begin{equation}
k = (1,0,1) \, .
\end{equation}
If no events are detected by $A$, then the 90\% confidence upper limit is given by (\ref{eqn:spul0}) with $\epsilon\to\eA+\eAB$: 
\begin{equation}\label{eqn:ulSIN0b}
\lambda_{90\%} = \frac{2.30}{\eA+\eAB} \, .
\end{equation}
If one event is detected by $A$, the limit is 
\begin{equation}\label{eqn:ulSIN1b}
\lambda_{90\%} = \frac{3.89}{\eA+\eAB} \, .
\end{equation}

The efficiency and background of the SINGLE combination are intermediate between those of the OR and AND combinations.  In general, $\eTOT \ge \eA+\eAB \ge \eAB$, so for a given number of measured events (for example, 0), OR will give the strongest limit, AND the weakest, and SINGLE an intermediate value.  On the other hand, the background is highest for OR and lowest for AND, so there is a greater chance of having $n>0$ events in the OR combination.  Unfortunately, for an unbiased analysis one must choose the upper limit method {\em before} counting events, so it is difficult to make the best choice between the AND, OR, and SINGLE options {\em a priori}. 

%The AND combination gives superior upper limits to the OR combination when there are events detected by some but not all pipelines.  (This will be likely, for example, if the pipelines have different responses to the background noise.)  It also tends to perform better for large-amplitude signals, when one expects all pipelines to have good detection efficiency and so $\epsilon_q\to\eTOT$.  The OR combination gives better upper limits when there are events detected by all of the pipelines.  It also tends to perform well at low signal amplitude, where the AND pipeline is limited by the least sensitive detector and $\epsilon_q\ll\eTOT$.  

\subsubsection{Efficiency-weighted combination} 
\label{sec:eff}

The AND, OR, and SINGLE options are just three examples of how one may select the exclusion surfaces for the multiple-pipeline counting experiment.  As just discussed, the relative strength of the upper limits one can achieve with these options depends on the number of events detected by each pipeline combination, which one does not know {\em a priori} in a blind analysis.

An obvious drawback of the AND and OR examples is that the exclusion surfaces are selected without regard to the known sensitivities $\vec{\epsilon}$ of the various pipeline combinations.  One expects that the strongest upper limits should involve use of this information.  As a trivial example, a pipeline combination with zero detection probability ($\epsilon_i=0$) should be ignored when setting upper limits ($n_i$ should be ignored).  The SINGLE combination makes some limited use of the known sensitivities, but throws away all of the information produced by the less-sensitive pipelines, even if they are only slightly less sensitive than the best one.

A more natural way to incorporate the efficiency information in the upper limit procedure is to orient the exclusion surfaces according to the measured efficiencies.  For plane exclusion surfaces, the simplest choice is 
\begin{equation}
\vec{k} = \vec{\epsilon} \, .
\end{equation}
We term this choice the {\em efficiency weighted combination}, or EFF.

Heuristically, the efficiency weighted combination is an intelligent 
choice because it places the largest emphasis on the measurements 
made by the most sensitive combinations of pipelines.  To see one of 
the desirable properties of this choice, consider a repeated experiment. 
%with low background ($\vec{b}\simeq0$).  
In an ensemble of experiments, the expected number of detections by each 
pipeline combination $i$ is %proportional to $\epsilon_i$; specifically,
\begin{equation}\label{eqn:exp}
\langle \vec{n} \rangle = \lambda_{\mathrm{true}} \vec{\epsilon} + \vec{b} \, .
\end{equation}
%For repeated experiments with low background it follows that
%\begin{equation}\label{eqn:exp2}
%\lambda_{\mathrm{true}}
%   =  \frac{
%        \langle \vec{n} \rangle \cdot \vec{\epsilon}
%      }{
%        \vec{\epsilon} \cdot \vec{\epsilon}
%      } \, .
%\end{equation}
Suppose the observed number of events is $\vec{n}'$ in one experiment, and $\vec{n}''$ in a second.  Which measurement should give the higher upper limit?  If $(\vec{n}''-\vec{n}')\cdot\vec{\epsilon} > 0$, then the second measurement is consistent with a higher limit on $\lambda$.  If $(\vec{n}''-\vec{n}')\cdot\vec{\epsilon} = 0$, then the two measurements imply the same upper limit on $\lambda$.  The choice $\vec{k}$ = $\vec{\epsilon}$ for the exclusion contours enforces these requirements.

\subsection{Example: Rate Limit vs.~Amplitude}
\label{sec:example}

Consider once more the case of two pipelines $A$ and $B$.  Let us suppose that the target signals are characterized by an amplitude $\rho$, and that the detection efficiencies of $A$ and $B$ separately, $E_A=\eA+\eAB$, $E_B=\eB+\eAB$, and their logical combinations $\eA,\eB,\eAB$, are as shown in Figure 1.  This scenario is typical of searches for gravitational-wave bursts by LIGO and similar detectors \cite{Ab_etal:04,Ab_etal:05,Ab_etal:05b,Ab_etal:07,S5firstyear}.  Our objective is to set an upper limit on $\lambda$ as a function of the signal amplitude $\rho$.

Since both $E_A$ and $E_B \to 1$ at large $\rho$, $\eAB\to 1$ as well, while $\eA$ and $\eB$ are nonzero for only a limited range of signal amplitudes.  In this toy model, $A$ is sensitive to slightly weaker signals than $B$, so $\eA>\eB$.  However, since both $\eA$ and $\eB$ are nonzero, each pipeline is able to detect some signals that the other pipeline misses.  Therefore, one expects that combining the measurements of the two pipelines should be able to provide more information on the event rate than either pipeline alone.

\begin{figure}
\begin{center}
  \includegraphics[width=0.75\textwidth]{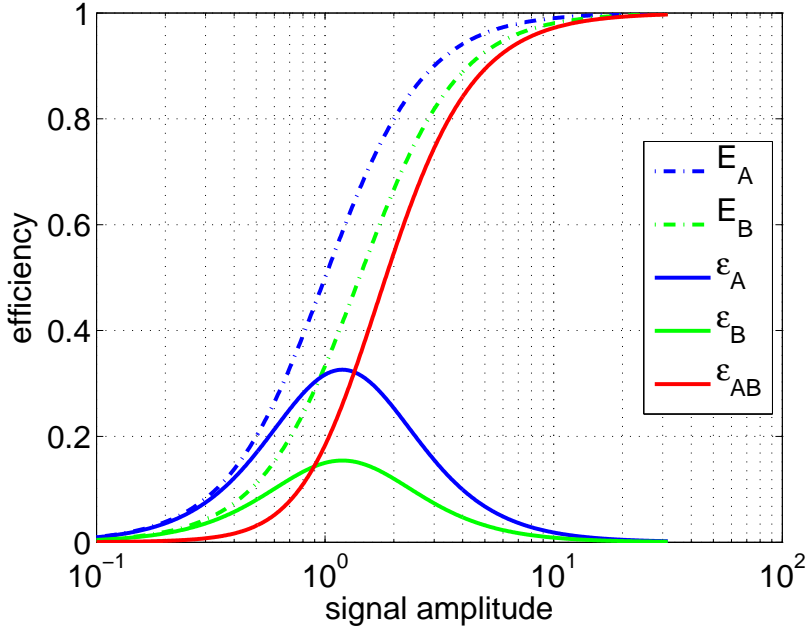}
  \caption{\label{fig:eff} Efficiencies for two pipelines $A$ and $B$.  In our toy model, the signal is characterized by an amplitude $\rho$.  The dotted lines $E_A=\eA+\eAB, E_B=\eB+\eAB$ show the efficiencies of the two pipelines considered separately.  The continuous lines show the efficiencies of the logical combinations of the pipelines: $\eA$ ($A$ not $B$), $\eB$ ($B$ not $A$), and $\eAB$ ($A$ and $B$).}
\end{center}
\end{figure}

Let us now compare the performance of four different choices of 
exclusion surfaces: AND, OR, SINGLE, and EFF.  For the moment, let 
us ignore any background when computing the upper limits; 
{\em i.e.,} we will use $\vec{b}=0$.  (We will compare limits 
including background in the next section.)

\begin{figure}
\begin{center}
  \includegraphics[width=0.75\textwidth]{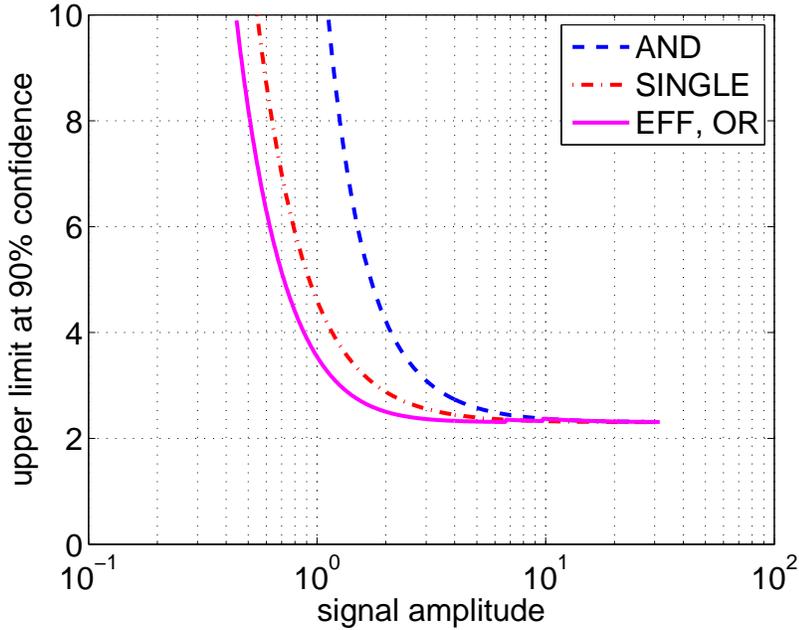}
  \caption{\label{fig:ul000} Upper limits as a function of signal amplitude when no events are detected.  All methods give the asymptotic limit $2.3$ for large amplitudes.  The EFF and OR combinations give the strongest limits at low amplitude because they have better detection efficiency than the AND and SINGLE combinations.}
\end{center}
\end{figure}

Consider first the case where no events are detected.  The upper 
limits from each combination are shown in Figure~\ref{fig:ul000}.  
All combinations give $\lambda_{90\%}=2.3$ at high amplitude, 
where $\eAB\to1$.  In particular, the EFF upper limit is
\begin{eqnarray}\label{eqn:ulEFF0a}
0.1 
  & = &  C_{\vec{k}}((0,0,0)|\lambda_{90\%}\vec{\epsilon}) \nonumber \\
  & = &  \e^{-\eTOT\lambda_{90\%}} \, ,
\end{eqnarray}
\begin{equation}\label{eqn:ulEFF0b}
\lambda_{90\%} = \frac{2.30}{\eTOT} \, ,
\end{equation}
identical to the OR limit.
The EFF and OR combinations give the strongest limits for weak 
signals because of their better efficiency (which is $\eTOT$, 
the sum of the efficiencies of all pipeline combinations).

\begin{figure}
\begin{center}
  \includegraphics[width=0.75\textwidth]{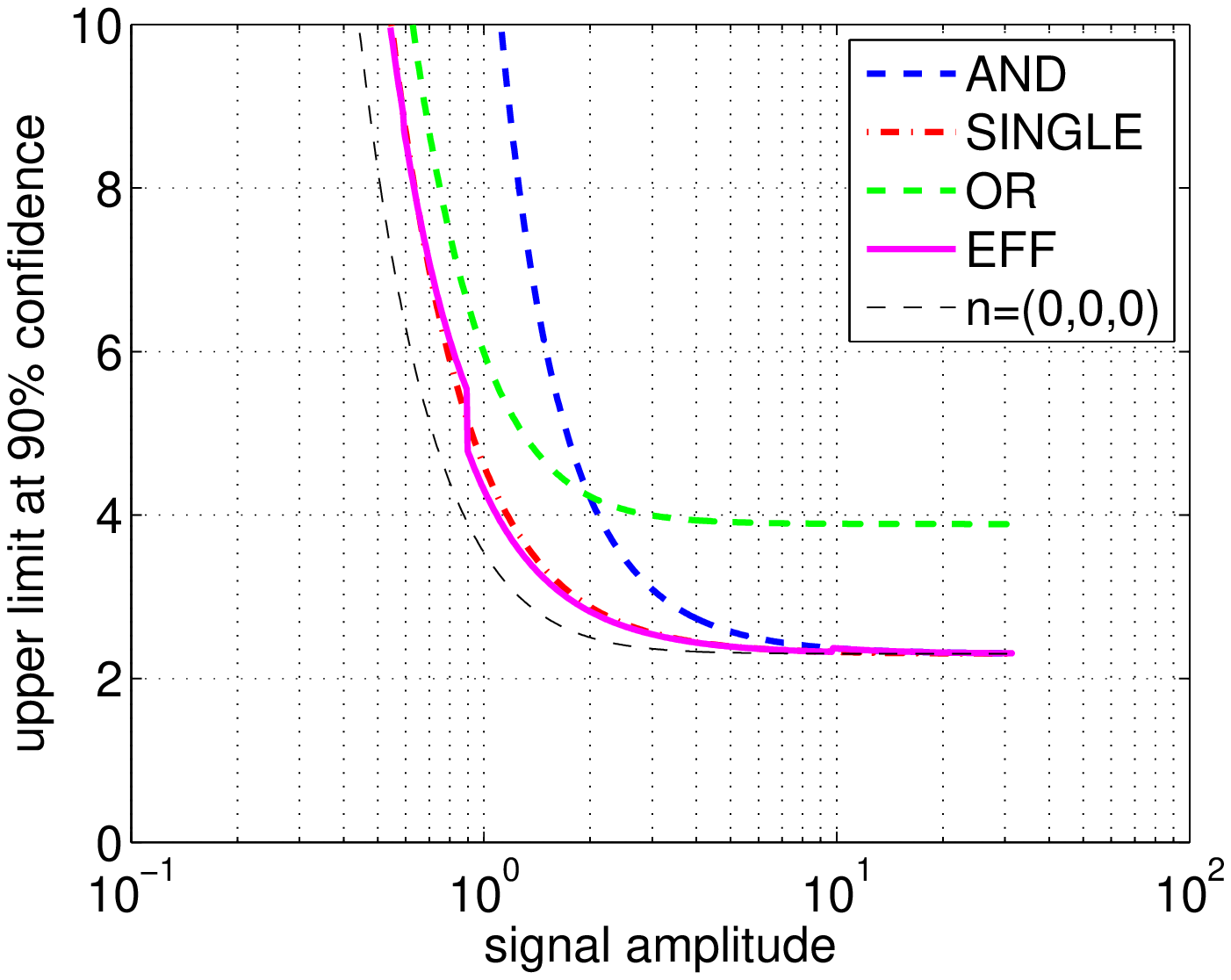}
  \caption{\label{fig:ul010} 
Upper limits as a function of signal amplitude when one event is detected by the less sensitive pipeline ($B$).  The OR combination asymptotes to the single-event value 3.9.  The lone event is not counted by the AND, SINGLE combinations, which give the $n=0$ limit 2.3.  The EFF combination ignores the event at high amplitudes (where $\eB\to0$), while at lower amplitudes the EFF limit is very close to the SINGLE limit as $\eB\ll\eA$.  The thin dashed line is the best possible upper limit from the counting experiment: that for zero observed events using the EFF or OR combinations (see Figure~\ref{fig:ul000}).}
\end{center}
\end{figure}

Now consider the case of one event detected by the weaker pipeline $B$: $\vec{n}=(0,1,0)$.  
The upper limits are shown in Figure~\ref{fig:ul010}.  The OR combination does poorly at high amplitudes because of the detected event.  The AND limit is much better at high amplitudes because $A$ did not see the event, but still poor at low amplitudes because $\eAB\to0$. The SINGLE combination performs well, giving the same result as the $n=0$ case, because it ignores the event counted by the less sensitive pipeline.  
The EFF upper limit is computed by summing over 
\begin{equation}
\vec{N}\cdot\vec{\epsilon} \le \vec{n}\cdot\vec{\epsilon} = \eB \, .
\end{equation}
For signal amplitudes $\rho\gtrsim1$, $\eB$ is the smallest efficiency, so the allowed terms are $\vec{N}\in\{(0,0,0),(0,1,0)\}$.  The EFF limit is then given by 
\begin{eqnarray}\label{eqn:ulEFF1a}
0.1 
  & = &  P((0,0,0)|\lambda_{90\%}\vec{\epsilon}) 
             + P((0,1,0)|\lambda_{90\%}\vec{\epsilon}) 
             \nonumber \\
  & = &  (1 + \eB\lambda_{90\%}) \e^{-\eTOT\lambda_{90\%}} \, .
\end{eqnarray}
The extra $\eB\lambda$ term makes the EFF upper limit only slightly higher than the $2.3/\eTOT$ value obtained in the $n=0$ case, as can be seen from Figure~\ref{fig:ul010}.  
For $\rho\lesssim1$, $\eB > \eAB$ and the upper limit includes additional $(\eAB\lambda)^{N_{AB}}$ terms.  This causes the EFF limit to increase, but again only slightly, as $\eAB$ is typically much smaller than $\eA$, $\eB$ at these low amplitudes.

We see that the EFF combination effectively ignores the event counted by the insensitive pipeline combination $B$, and gives a limit as good as or even slightly better than that from the SINGLE combination.

\begin{figure}
\begin{center}
  \includegraphics[width=0.75\textwidth]{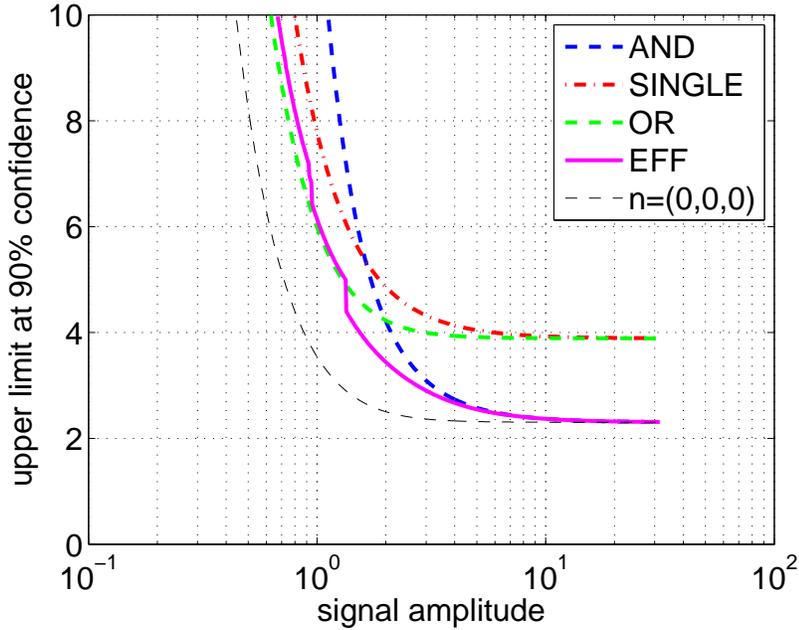}
  \caption{\label{fig:ul100} 
Upper limits as a function of signal amplitude when one event is detected by the more sensitive pipeline ($A$).  The OR and SINGLE combinations asymptote to the single-event value 3.9.  The lone event is not counted by the AND combination, which gives the $n=0$ limit 2.3.  The EFF combination ignores the event at high amplitudes (where $\eA\to0$), while at lower amplitudes the EFF limit is very close to the OR limit for $n=1$.  The thin dashed line is the best possible upper limit from the counting experiment: that for zero observed events using the EFF or OR combinations (see Figure~\ref{fig:ul000}).}
\end{center}
\end{figure}

Now turn to the case in which one event is detected by the more sensitive pipeline, A: $\vec{n}=(1,0,0)$.  
The upper limits are shown in Figure~\ref{fig:ul100}.  Again, the OR combination does poorly at high amplitudes because of the detected event.  
The SINGLE combination does even worse, since the event was found by the more sensitive pipeline, and the SINGLE combination has lower efficiency than the OR combination.  The AND combination again performs well at high amplitudes and poorly at low amplitudes. 
The EFF upper limit is computed by summing over 
\begin{equation}
\vec{N}\cdot\vec{\epsilon} \le \vec{n}\cdot\vec{\epsilon} = \eA \, .
\end{equation}
The number of terms in the sum depends on the relative values of $\eA$, $\eB$, and $\eAB$.  In this simple example, for $\rho>1.4$, $\eA=2\eB<\eAB$ and the allowed terms are $\vec{N}\in\{(0,0,0),(1,0,0),(0,1,0),(0,2,0)\}$.  The EFF limit is given by 
\begin{eqnarray}\label{eqn:ulEFF2a}
0.1 
  & = &  (1 + \eA\lambda_{90\%} + \eB\lambda_{90\%} 
%  \nonumber \\
%  &   &  \mbox{}
  + \eB^2\lambda_{90\%}^2) \e^{-\eTOT\lambda_{90\%}} \, .
\end{eqnarray}
Since $\eA$ and $\eB$ are small at high amplitudes, the upper limit is 
again similar to the $n=0$ value of $2.3/\eTOT$.  For $\rho<1.4$, 
$\eA > \eAB$ and the cumulative distribution 
$C_{\vec{k}}(\vec{n}|\lambda_\alpha\vec{\epsilon})$ in (\ref{eqn:ulk})
includes additional $(\eAB\lambda)^{N_{AB}}$ terms.  This causes 
the EFF limit to increase, becoming similar to that from the OR 
combination.  In short, the EFF combination gives the strongest 
limits at high amplitudes because pipeline $B$ should have seen 
a real event there and did not, and it gives the strongest limits at 
low amplitudes because it has better efficiency than the AND 
combination.

\begin{figure}
\begin{center}
  \includegraphics[width=0.75\textwidth]{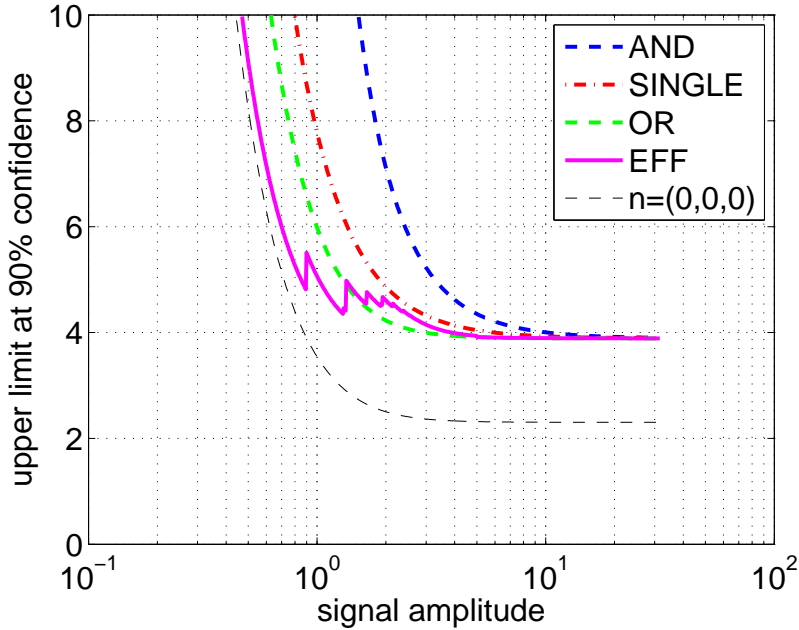}
  \caption{\label{fig:ul001} 
Upper limits as a function of signal amplitude when a single event is detected by both pipelines ($A$ and $B$).  All combinations give the asymptotic limit of 3.9 at large amplitudes.  The EFF combination ignores this event at low amplitudes (where $\eAB\ll\eA,\eB$) and tends to the zero-event limit for $\rho<1$.
The thin dashed line is the best possible upper limit from the counting experiment: that for zero observed events using the EFF or OR combinations (see Figure~\ref{fig:ul000}).
}
\end{center}
\end{figure}

Finally, consider the case of a single event detected by both pipelines: $\vec{n}=(0,0,1)$.  In this case all combinations give the asymptotic limit of 3.9 at large amplitudes, as seen in Figure~\ref{fig:ul001}.  The relative limits of the AND, OR, and SINGLE combinations are the same as in the $n=0$ case.  We see, however, that the EFF combination outperforms all other combinations (including OR) in the low-amplitude limit.  In fact, the EFF limit reaches nearly the $n=0$ value at low signal amplitudes.  This counter-intuitive result has a simple explanation: at low amplitudes ($\rho<1$), the probability $\eAB$ of a real event being detected jointly by $A$ and $B$ is much smaller than the probabilities $\eA$, $\eB$ of it being detected by either pipeline alone.  The observation $\nAB>0$ is therefore inconsistent with the hypothesis of a low-amplitude signal.  The efficiency weighted combination therefore ignores this measurement for the low-amplitude upper limits, and the limit is dominated by the measurements $n_A=0=n_B$.  

It is worth noting that the upper limits obtained from the efficiency-weighted procedure are neither monotonic nor continuous; this is most evident in Figure~\ref{fig:ul001}.  The limits are not monotonic because the efficiencies $\eA$, $\eB$, $\eAB$ of the logical combinations of pipelines are not monotonic, as shown in Figure~\ref{fig:eff}.  The origin of the discontinuities is slightly more subtle; it arises from the need to sum over a discrete set of $\vec{N}$ in (\ref{eqn:ulk}).  For the efficiency-weighted combination, the condition (\ref{eqn:contour}) depends on the assumed signal amplitude through the efficiencies, $\vec{k}=\vec{\epsilon}(\rho)$.  Therefore, the sum may include different numbers of terms for different signal amplitudes.  The discontinuities occur at signal amplitudes where another term satisfies the condition to be included in the sum in (\ref{eqn:ulk}), 
$\vec{N}\cdot\vec{\epsilon} \le \vec{n}\cdot\vec{\epsilon}$.
In turn, this happens when the ratio of efficiencies equals a rational number.
We stress that these discontinuities are a general feature of using efficiencies to weight the pipeline combinations, and that they are not indicative of any problem with the procedure.  The upper limits at different $\rho$ values are limits on different signal models, and therefore they need not be continuous or monotonic functions of $\rho$.  Indeed, this behaviour is advantageous, as seen in Figure~\ref{fig:ul001}, where the efficiency-weighted upper limit is able to drop below the OR limit at low amplitudes.

In each of the cases considered, efficiency weighting gives upper limits as approximately as strong as or stronger than any of the other choices.  Without efficiency weighting, the best remaining combination is different for the different cases: AND, OR, and SINGLE each perform best for at least one of the cases tested.  While we must chose the weighting 
{\em before} measuring $\vec{n}$ for the upper limit procedure to have the proper coverage, there is no way to know {\it a priori} whether to choose AND, OR, or SINGLE.  The efficiency-weighted combination, however, gives optimal or near-optimal performance in all cases.

We can gain insight into the strong performance of the efficiency weighting choice by examining the form of the upper limit equation (\ref{eqn:ulk}):
\begin{equation}
1-\alpha = (1+\epsilon_1\lambda_\alpha+\frac{\epsilon_1^2\lambda_\alpha^2}{2}+\ldots+\epsilon_2\lambda_\alpha+\ldots)\e^{-(\epsilon_1 + \ldots)\lambda_\alpha} \, .
\end{equation}
The set of efficiencies $\epsilon_i$ appearing in the exponential is determined by the choice of pipeline combination used for the upper limit.  The set of $\epsilon_i$ terms  appearing in the factor in front of the exponential depends on the set of measured events $\vec{n}$ as well as the pipeline combination chosen.  
As a rule, adding efficiency terms in the exponential {\em decreases} the upper limit.  Adding efficiency terms to the factor in front of the exponential {\em increases} the upper limit.  
For the AND and SINGLE combinations, only some of the efficiencies appear in the exponential.  
With the OR and EFF combinations, the efficiencies for all pipeline combinations appear in the exponential, giving the maximum efficiency possible ($\eTOT$).  Between these two, the EFF combination will typically give fewer terms in the prefactor when events are detected with the less sensitive pipeline combinations.  This will result in a lower limit than the OR combination.  It may have more terms when the most sensitive combination sees the event, thus giving a higher limit than the OR combination in these cases.  As seen in Figure~\ref{fig:ul001}, this loss in upper limit tends to be small; since the extra terms are associated with low-efficiency pipeline combinations, and appear with powers of those small $\epsilon_i$.

\subsection{Upper Limits with Background}
\label{sec:background}

We have seen that the EFF weighted combination tends to give stronger 
upper limits than the AND, OR, and SINGLE weightings when we ignore 
the background.  We now demonstrate by example that this superior 
performance continues when we account for the background as well.
We do this by computing the expectation value of the upper limit 
as a function of the true foreground rate $\lambda$ for two scenarios: 
one with low background, and one with high background.

Let us consider once more the case of our two pipelines $A$ and $B$.  
We will work initially with a fixed set of efficiencies, 
\begin{equation}
\vec{\epsilon} = (\eA,\eB,\eAB) = (0.345,0.175,0.480) \, .
\end{equation}
Let us assume the background to be 
\begin{equation}
\vec{b} = (\bA,\bB,\bAB) = (1/3,1/3,1/3) \, \bTOT \, .
\end{equation}
With this background, on average, pipelines $A$ and $B$ detect the 
same number of background events, and half of the events detected 
by one are also detected by the other.  The total expected 
background is $\bTOT$.  We will consider the cases $\bTOT=0.1$ 
(``low background'') and $\bTOT=1$ (``high background'').  

\begin{figure}
\begin{center}
  \includegraphics[width=0.75\textwidth]{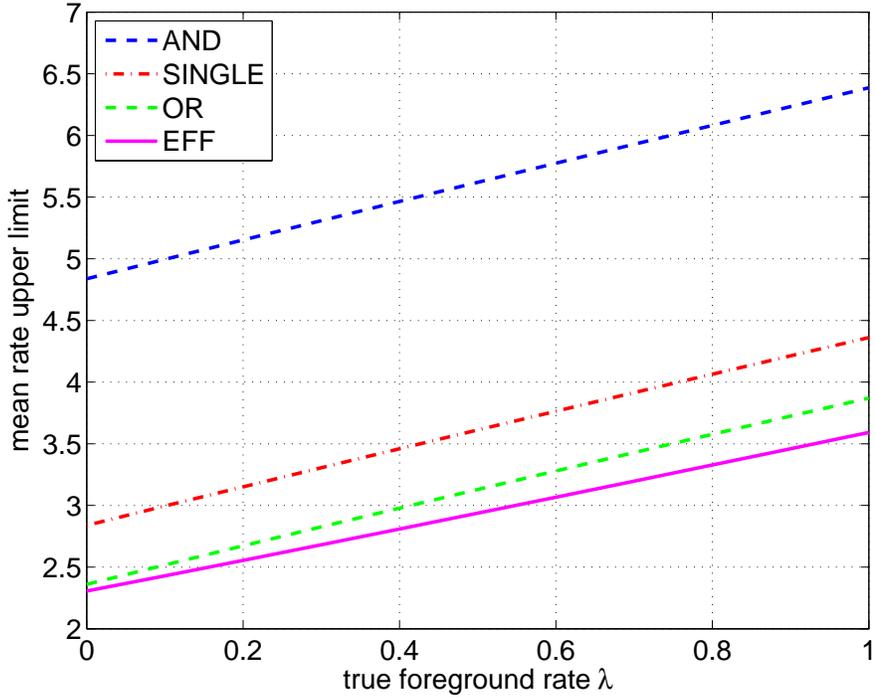}
  \caption{\label{fig:lowbckgrd} 
Mean upper limit as a function of the true foreground rate $\lambda$ 
in an ensemble of experiments with fixed low background.  This 
two-pipeline experiment has efficiency 
$(\eA,\eB,\eAB) = (0.345,0.175,0.480)$ and background 
$(\bA,\bB,\bAB) = (1/30,1/30,1/30)$.
}
\end{center}
\end{figure}

\begin{figure}
\begin{center}
  \includegraphics[width=0.75\textwidth]{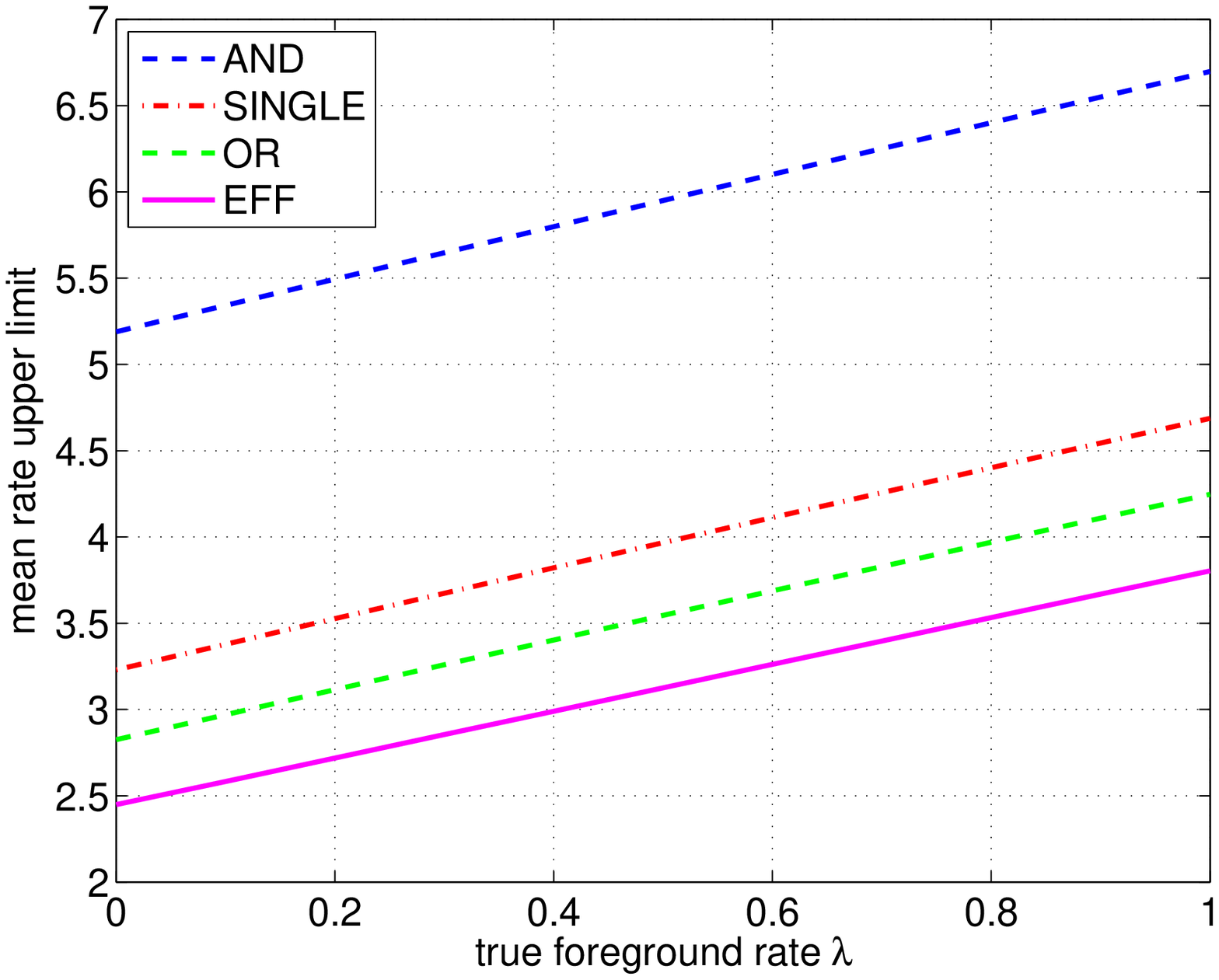}
  \caption{\label{fig:highbckgrd} 
Mean upper limit as a function of the true foreground rate $\lambda$ 
in an ensemble of experiments with fixed high background.  This 
two-pipeline experiment has efficiency 
$(\eA,\eB,\eAB) = (0.345,0.175,0.480)$ and background 
$(\bA,\bB,\bAB) = (1/3,1/3,1/3)$.
}
\end{center}
\end{figure}

A straightforward Monte Carlo analysis was used to estimate the 
upper limit in an ensemble of experiments.  Figure \ref{fig:lowbckgrd} 
shows the mean limits from the AND, OR, SINGLE, and EFF combinations 
as a function of the true value of $\lambda\in[0,1]$ for the low background 
case.  Figure \ref{fig:highbckgrd} shows the mean limits for the high 
background case.  In both cases the EFF weighting gives stronger limits 
than any of the other weightings for all values of $\lambda$ tested.  
The gap between the EFF upper limits 
and the next best limits (from OR) is particularly large for the 
high-background case.  These findings support our conclusion that 
the EFF weighting ``protects'' the upper limit against modest
background contamination.

\begin{figure}
\begin{center}
  \includegraphics[width=0.75\textwidth]{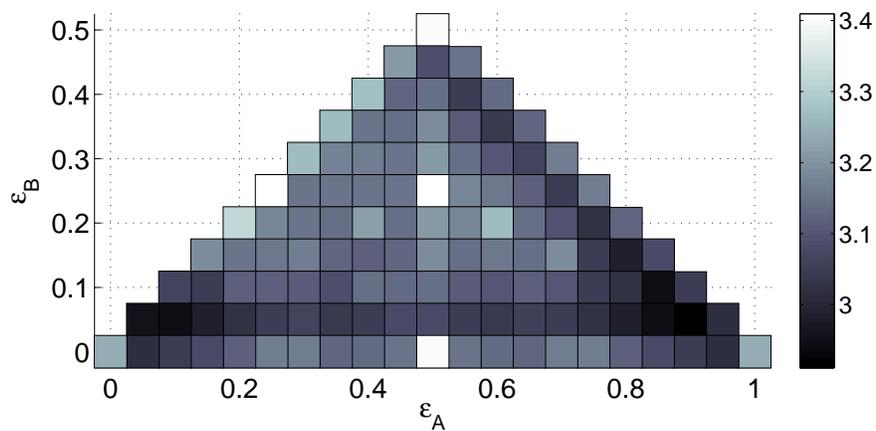}
  \caption{\label{fig:UL2D} 
Mean upper limit as a function of efficiency $\vec{\epsilon}=(\eA,\eB,1-\eA-\eB)$  
in an ensemble of experiments with background 
$(\bA,\bB,\bAB) = (1/3,1/3,1/3)$ and true event rate $\lambda=0.5$. 
The largest limits occur when $\eA$, $\eB$, and $\eAB=1-\eA-\eB$ 
are related by the ratio of small integers, as discussed in 
Section~\ref{sec:example}.
}
\end{center}
\end{figure}

To get a sense of the robustness of the EFF weighting performance, we 
repeat the Monte Carlo for a range of efficiencies.  Specifically, we vary 
$\eA$ over [0,1], $\eB\le\eA$, and keep $\eAB=1-\eA-\eB$ so that $\eTOT=1$.  
We use $\bTOT=1$ (``high background'') and $\lambda=0.5$.
Figure~\ref{fig:UL2D} shows how the mean upper limit from the EFF weighting 
varies with $\eA$, $\eB$.  The mean limits range from 2.91 to 3.41, a variation
of less than 20\%.  By contrast, the mean limits from the other weightings 
(not shown) are always higher: $\ge 3.40$ (SINGLE); $\ge 3.25$ (AND), and $=3.55$ (OR).
This indicates that the superior performance of the EFF weighting 
is not reliant on the efficiencies taking particular values.

It can be noted from Figure~\ref{fig:UL2D} that the EFF limit  
does not reduce to the SINGLE limit ($\ge3.40$) when $\eB\to0$.  
This is because the EFF combination becomes 
$\vec{k}=(\eA,0,\eAB)$, 
whereas the SINGLE weighting is $\vec{k}=(1,0,1)$.  
So, the EFF weighting maintains a distinction between events 
detected by A alone and those detected jointly by A and B.
The result is that the EFF limits are lower than or equal to the SINGLE 
limits as $\eB\to0$, with equality at $\vec{\epsilon}=(0.5,0,0.5)$.

Finally we note that the EFF weighting, since it is based on efficiency alone, 
is most applicable to the case where the background is relatively small.  We 
concentrate on the case where the expected number of events due to 
background of order 1 or less.  For much higher backgrounds the 
optimal weightings should also include information on the backgrounds $\bA$, 
$\bAB$, \ldots of the various pipeline combinations.

\section{Multiple Data Sets}
\label{sec:data}

The formalism we have developed for multiple algorithms analyzing a common data set can be applied equally well to the analysis of multiple sets of data.  For example, we may have data from several observation periods, each characterized by the use of a different set of instruments, or over which the sensitivity of the instruments changed, {\em etc}.
In this case, the analyses of the separate data epochs may be considered as separate pipelines for purposes of setting an upper limit.

As a simple example, consider the case of a single algorithm used to 
analyse data from two disjoint data sets $A$ and $B$, with durations 
$T_A$, $T_B$.  The sensitivity of the experiment is characterized by 
the two numbers 
\begin{itemize}
\item[$\eA$:]  The probability that any given foreground event will be 
detected during period $A$;
\item[$\eB$:]  The probability that any given foreground event will be 
detected during period $B$.
\end{itemize}
The background is characterized by 
\begin{itemize}
\item[$\bA$:]  The expected number of background events detected 
during period $A$;
\item[$\bB$:]  The expected number of background events detected 
during period $B$.
\end{itemize}
The outcome of the experiment is the set of two numbers
\begin{itemize}
\item[$\nA$:]  The number of events detected during period $A$;
\item[$\nB$:]  The number of events detected during period $B$.
\end{itemize}
Since any given event can be detected during period $A$ or period 
$B$ but not both, we have $\eAB=0$, $\bAB=0$, $\nAB=0$.  We see 
immediately that this is a special case of the two-pipeline analysis, 
where we treat the analysis of the separate data sets as separate 
pipeline measurements.  In fact, it is a particularly simple case, 
as we know $\eAB=0$, $\bAB=0$, $\nAB=0$ {\em a priori}.

Note that we define the efficiencies $\eA$, $\eB$ in terms of the probability of events from anywhere in the {\em entire} observation period $T$ being detected during periods $A$ or $B$.  We are taking the union of the data sets to treat them as one large set.  This is the most convenient approach, since it matches precisely how the multiple-pipeline case was developed.  It saves us from including the separate observation times $T_A$, $T_B$ explicitly in our upper limit calculations.  Instead, they are included implicitly in the efficiencies.  For example, $\eA$ has a maximum possible value of $T_A/(T_A+T_B)$.

For concreteness, let us suppose we have two data sets of equal 
length, $T_A = T_B = 0.5 T$.  Suppose also that the instruments 
used were more sensitive during period $A$, such that $\eA=3/5$, 
$\eB=2/5$, and $\eTOT=\eA+\eB=1$.  Table~\ref{tab:ULs1} shows 
the upper limits obtained ignoring the background.  We compare 
the OR (combining event counts from both periods), SINGLE (only 
counting events from the more sensitive period), and EFF 
combinations for zero or one detected event.  (The AND combination 
is not applicable to this case, since $\eAB=0$.)

In each case, the EFF combination gives the best upper limit. For 
no detected events, the EFF and OR combinations give the limit 2.3 
as before.  The SINGLE limit is a factor 5/3 higher, because it 
uses only 3/5 of the integrated sensitivity of the experiment 
($\eA=3\eTOT/5$).  For one event detected in the less sensitive 
period $B$, the EFF combination gives the best upper limit -- even 
better than SINGLE.  This may be surprising, in that the SINGLE 
upper limit is computed for zero events.  We see that the extra 
sensitivity gained by including the $B$ measurement in the EFF 
upper limit more than offsets the loss in the limit due to having 
a detected event.  Finally, for the case of one event detected in 
the more sensitive period $A$, the EFF limit matches the OR limit.  Interestingly enough, the SINGLE combination performs worse than 
EFF in all cases; for the given efficiencies, we always get a 
better limit by using all of the data.

\begin{table}
\caption{\label{tab:ULs1} 
Comparison of upper limits obtained for various possible outcomes of 
a counting experiment on two data sets $A$ and $B$ with $\eA=3/5$, 
$\eB=2/5$, and ignoring background.  The cases are: no events detected 
($\vec{n}=(0,0,0)$); one event detected in B ($\vec{n}=(0,1,0)$); one 
event detected in A ($\vec{n}=(1,0,0)$).}
\begin{indented}
\item[]\begin{tabular}{cccc}
\br
          & \multicolumn{3}{c}{upper limit} \\ \cline{2-4}
$\vec{n}$ & OR  & SINGLE & EFF \\
\mr
(0,0,0)   & 2.3 &   3.8  & 2.3 \\
(0,1,0)   & 3.9 &   3.8  & 3.1 \\
(1,0,0)   & 3.9 &   6.5  & 3.9 \\
\br
\end{tabular}
\end{indented}
\end{table}

For a larger difference in efficiencies, the differences in upper 
limits are more pronounced.  Table~\ref{tab:ULs2} compares the 
upper limits for $\eA=2/3$, $\eB=1/3$, $\eTOT=1$.  The OR limits 
are unchanged.  The SINGLE limits are better than those in 
Table~\ref{tab:ULs1} because the SINGLE combination now contains 
2/3 of the integrated sensitivity of the experiment ($\eA=2\eTOT/3$) 
instead of only 3/5.  The changes in the EFF limits are more 
complicated.  For one event detected in the less sensitive period 
$B$, the EFF combination still gives the best upper limit -- 
slightly better than before, because the weighting of $B$ is less 
than in the previous case.  For one event detected in $A$, the EFF 
limit is between the OR limit and the SINGLE limit.  The increase 
over the limit in Table~\ref{tab:ULs1} is due to the fact that for 
$\eA=2\eB$, the cumulative sum in (\ref{eqn:ulk}) now includes the 
terms $\vec{N}=\{(0,0,0),(1,0,0),(0,1,0),(0,2,0)\}$, whereas for 
$\eA=1.5\eB$ it includes only $\vec{N}=\{(0,0,0),(1,0,0),(0,1,0)\}$.

Note that the EFF limits are particularly robust against background events contaminating the less-sensitive data sets.  This allows the sub-optimal data to be used to strengthen scientific results without fear of ``spoiling'' the upper limits.
In particular, note that the average of the upper limits for the single-event cases ($\vec{n}=(1,0,0)$ and $(0,1,0)$) is best for the EFF combination in both Table~\ref{tab:ULs1} and Table~\ref{tab:ULs2}.  So, if the data sets have equal background probability (assumed $\ll1$), the EFF combination will on average give the best upper limits for low true event rates.

\begin{table}
\caption{\label{tab:ULs2} 
Comparison of upper limits obtained for various possible outcomes of 
a counting experiment on two data sets $A$ and $B$ with $\eA=2/3$, 
$\eB=1/3$, and ignoring background.  The cases are: no events detected 
($\vec{n}=(0,0,0)$); one event detected in B ($\vec{n}=(0,1,0)$); one 
event detected in A ($\vec{n}=(1,0,0)$).}
\begin{indented}
\item[]\begin{tabular}{cccc}
\br
          & \multicolumn{3}{c}{upper limit} \\ \cline{2-4}
$\vec{n}$ & OR  & SINGLE & EFF \\
\mr
(0,0,0)   & 2.3 &   3.5  & 2.3 \\
(0,1,0)   & 3.9 &   3.5  & 3.0 \\
(1,0,0)   & 3.9 &   5.8  & 4.3 \\
\br
\end{tabular}
\end{indented}
\end{table}

Finally, since we have seen benefits from treating multiple data sets separately, one might ask if we should always split up data sets.  In particular, why not sub-divide all data sets {\em ad infinitum}?  The answer comes from noting that the benefits of the EFF combination arise from exploiting differences in the efficiencies $\epsilon_i$.  If the differences in efficiency between two data sets are negligible, then there is no benefit to treating them separately.  
For example, for two sets of data with identical efficiencies, the EFF and OR combinations will always give identical limits: since $\nAB=0$ always, choosing $\vec{k}=\vec{\epsilon}$ will always give the same results as $\vec{k}=(1,\ldots,1)$.  One therefore gets no benefit from sub-dividing epochs of constant sensitivity.

\section{Summary}
\label{sec:summary}

We have proposed a general technique for setting upper limits on 
Poisson processes from counting experiments involving multiple data 
sets and multiple event-counting algorithms (which we collectively 
refer to as multiple ``pipelines'').  This technique is an extension 
of the standard procedure for one-sided classical confidence intervals.  
There are two key features.  First, we characterize the measurements 
by the logical combinations of pipelines -- the number of events 
counted by A-and-B, by A-and-not-B, {\em etc.}  Second, we select a 
rank-ordering of the space of possible measurements which is based 
on the relative detection efficiencies of these logical combinations.  
This efficiency weighting uses all of the counts from the experiment, 
but assigns more significance to those counts from pipeline 
combinations which are expected to detect more foreground events.  
We have seen that in typical cases for low background and low 
foreground event rate, the efficiency weighting tends to give 
stronger upper limits than selecting the AND or OR combination 
of pipelines, or selecting the single most sensitive pipeline 
only.  In particular, the efficiency weighting procedure tends 
to be robust against modest background contamination of the event 
counts.  This allows all of the observational results to contribute 
to the upper limit while reducing the chances that background 
contamination of some counts will weaken it.

%EFF useful when different pipelines have different relative sensitivities 
%to signals of different type/energy/amplitude/etc.

In this paper we have focused on computing upper limits; however, 
the method has wider applicability.  The characterisation of the 
experiment in terms of logical combinations of pipelines and 
the subsequent rank ordering effectively reduce the space of 
measurements to one dimension.  At this point we are free to apply 
other standard procedures for constructing one- or two-sided confidence 
intervals.  It would be interesting, for example, to apply the 
Feldman-Cousins procedure \cite{FeCo:98} to produce unified upper 
limits and confidence intervals for our multiple-pipeline experiment; 
we leave this consideration to the future.

As a final note, let us point out that the concept of a ``pipeline'' is quite general -- it is nothing more than a way of defining a count of events.  We have seen that different pipelines may consist of different algorithms applied to the same data, or the same algorithm applied to different data sets.  Distinct pipelines may also be defined in other ways, such as by applying a single algorithm to a single data set and segregating the resulting events into groups by some other attribute.  %, such as measured frequency, duration, or energy.  
For example, in gravitational-wave burst searches the background is largely due to events detected at low frequencies ($<200$ Hz).  Dividing events into low-frequency ($<200$ Hz) and high-frequency ($>200$ Hz) sets would produce limits on high-frequency gravitational waves that are not compromised by the low-frequency background.
LIGO matched-filtering searches for gravitational waves from inspiralling 
binaries \cite{abbott:122001,abbott:047101} use a similar idea, dividing 
the space of templates (signal parameters) into several regions.  Background 
events that match templates in one region then have minimal impact on the 
limits set in other regions of the template/signal space.
Multiple applications of the same algorithm with different counting 
thresholds can also be treated as separate pipelines and handled by 
our method; this might be appropriate when low- and high-amplitude 
events are produced by separate populations.  A multiple-threshold 
approach would have the benefit that events detected with low (high) 
amplitude have minimal impact on the rate limits set on the high (low) 
amplitude population.  Our method even naturally handles the case of a 
``veto'' analysis, in which one pipeline (B) processes data in such a 
way as to be deliberately {\em insensitive} to signals, but sensitive to 
background noise: $\eB,\,\eAB\simeq0$ but $\bB,\,\bAB>0$.  
The EFF weighting then automatically ignores (vetoes) events detected 
by A that are also detected by the veto pipeline B.

\section*{Acknowlegements}

The author would like to thank Lindy Blackburn, Shourov Chatterji, 
Jolien Creighton, Stephen Fairhurst, and Erik Katsavounidis for 
stimulating discussions, and Patrick Brady for valuable feedback 
on an earlier draft.  
This work was supported in part by STFC grant PP/F001096/1.

\appendix

\section{Derivative of $C_{\cal S}(\vec{n}|\vec{\epsilon},\lambda)$}
\label{sec:proof}

In this appendix we prove equation (\ref{eqn:prop2n}), 
\begin{eqnarray}
\frac{\rmd C_{\cal S}(\vec{n}|\lambda\vec{\epsilon}+\vec{b})}{\rmd\lambda} & < & 0 \, , \label{eqn:prop2nA}
\end{eqnarray}
where $\vec{n}$, $\vec{\epsilon}$, $\vec{b}$, and the family 
${\cal S}(\zeta)$ are held fixed.

First, we recall the definition (\ref{eqn:C}) of the cumulative probability $C_{\cal S}$,\begin{eqnarray}\label{eqn:CA}
C_{\cal S}(\vec{n}|\lambda\vec{\epsilon}+\vec{b}) 
  & = &  \!\!\! \sum_{\vec{N}|\zeta(\vec{N}) \le \zeta(\vec{n})} 
         P(\vec{N}|\lambda\vec{\epsilon}+\vec{b}) 
            \nonumber \\
  & = &  \!\!\! \sum_{\vec{N}|\zeta(\vec{N}) \le \zeta(\vec{n})}  \prod_{i=1}^q P(N_i|\epsilon_i\lambda+b_i)  
            \nonumber \\
  & = &  \!\!\! \sum_{\vec{N}|\zeta(\vec{N}) \le \zeta(\vec{n})}  \prod_{i=1}^q \frac{(\epsilon_i\lambda+b_i)^{N_i}\e^{-\epsilon_i\lambda-b_i}}{N_i!}   \, . \qquad
\end{eqnarray}
Taking the derivative with respect to $\lambda$ yields
%\begin{widetext}
\begin{eqnarray}\label{eqn:deriv}
\fl
\frac{\rmd C_{\cal S}(\vec{n}|\lambda\vec{\epsilon}+\vec{b})}{\rmd\lambda} 
  & = &  \sum_{\vec{N}|\zeta(\vec{N}) \le \zeta(\vec{n})}  \left[ 
             \frac{N_1 \epsilon_1 (\epsilon_1\lambda+b_1)^{N_1-1}}{N_1!} 
             \times \ldots \times
             \frac{(\epsilon_q\lambda+b_q)^{N_q}}{N_q!} 
             + \ldots 
             \right. \nonumber \\
  &   &      \left. \mbox{}
             + \frac{(\epsilon_1\lambda+b_1)^{N_1}}{N_1!} 
             \times \ldots \times
             \frac{N_q \epsilon_q (\epsilon_q\lambda+b_q)^{N_q-1}}{N_q!} 
             \right. \nonumber \\
  &   &      \left. \mbox{}
             - \left(\epsilon_1+\ldots+\epsilon_q\right) 
             \frac{(\epsilon_1\lambda+b_1)^{N_1}}{N_1!} 
             \times \ldots \times
             \frac{(\epsilon_q\lambda+b_q)^{N_q}}{N_q!} 
         \right]
         \nonumber \\
  &   &  \mbox{} \times
         \exp\left[-\left(\epsilon_1+\ldots+\epsilon_q\right)\lambda-(b_1+\ldots+b_q)\right]
         \, .
\end{eqnarray}
%\end{widetext}
Consider the contribution of the term $\vec{N}'=(N_1',\ldots,N_q')$ to the sum.  
We see that the positive terms arise from taking the derivative of the $\lambda^{N_i}$.  Each such positive term is exactly cancelled by a negative term coming from the derivative of the exponential from the term $\vec{N}''=(N_1',\ldots,N_i'-1,\ldots,N_q')$.  $\vec{N}''$ will always be included in the sum if $\vec{N}'$ is included because of the requirement that the normal to the surfaces ${\cal S}(\zeta)$ must have only non-negative components.  Therefore, all positive terms in (\ref{eqn:deriv}) are cancelled and the derivative must be negative.

\section*{References}
\bibliography{combinedul}
\bibliographystyle{unsrt}

\end{document}